\documentclass[12pt]{article}
\usepackage{amssymb,amsmath,epsfig,cite} 

\parskip=\medskipamount
\def\ba{\begin{eqnarray}}
\def\ea{\end{eqnarray}}

\def\b{\beta}                   
\def\g{\gamma}                  
\def\e{\epsilon}                

\def\wt{\widetilde}

\newcommand{\bea}{\begin{eqnarray}}
\newcommand{\eea}{\end{eqnarray}}

\font\tengoth=eufm10 \font\sevengoth=eufm7 \font\fivegoth=eufm5
\newfam\gothfam
\textfont\gothfam=\tengoth \scriptfont\gothfam=\sevengoth
  \scriptscriptfont\gothfam=\fivegoth

\begin{document}

\title{The confluent algorithm in second order \\ supersymmetric quantum
mechanics}

\author{David J. Fern\'andez C.${}^\dagger$ and  
Encarnaci\'on Salinas-Hern\'andez${}^\ddagger$ \\[8pt]
{\small ${}^\dagger$ Departamento de F\'{\i}sica, CINVESTAV} \\ 
{\small A.P. 14-740, 07000 M\'exico D.F., Mexico} \\[8pt]
{\small ${}^\ddagger$ Escuela Superior de F\'{\i}sica y Matem\'aticas and}
\\
{\small Escuela Superior de C\'omputo, Instituto Polit\'ecnico Nacional}
\\
{\small Ed. 9, U.P. Adolfo L\'opez Mateos, 07738 M\'exico D.F., Mexico}}
\date{}
\maketitle 

\begin{abstract} 
The confluent algorithm, a degenerate case of the second order
supersymmetric quantum mechanics, is studied. It is shown that the
transformation function must asymptotically vanish to induce non-singular
final potentials. The technique can be used to create a single level above
the initial ground state energy. The method is applied to the free
particle, one-soliton well and harmonic oscillator.
\end{abstract}

\section{Introduction}

The second order supersymmetric quantum mechanics (2-SUSY QM), which
involves second order differential intertwining operators
\cite{ais93,aicd95,sa96,fe97,fgn98,ro98,dnnr99,anst01}, has proved useful
to surpass the difficulty of `modifying' the excited state levels inherent
to the standard first order 1-SUSY QM. In fact, in the 2-SUSY treatment we
do not respect the restrictions imposed by 1-SUSY on the transformation
functions $u_1(x), \ u_2(x)$, i.e., they can have nodes but induce a
non-singular 2-SUSY transformation \cite{sa99,fnn00,fmrs02a,fmrs02b}. In
this way, potentials with two extra bound states above the ground state
energy of the initial Hamiltonian $H$ (or above the lowest band edge if
$V(x)$ is periodic) have been recently generated
\cite{sa99,fmrs02a,fmrs02b}.  A different atypical method employing two
complex conjugate factorization energies has been as well implemented,
generating in this case families of real isospectral 2-SUSY partner
potentials of $V(x)$ \cite{fmr02}.

There is yet another situation worth studying in detail, namely, when the
two factorization energies tend to a common {\it real} value $\epsilon$. 
This so-called {\it confluent algorithm} was used to generate a particular
family of isospectral oscillator potentials \cite{mnr00}. However, we have
not detected a generic analysis (for arbitrary factorization energies
$\epsilon$) of the properties of the transformation function $u(x)$
ensuring that the final potential will be non-singular. This is the
subject of the present paper, which has been organized as follows. In
section 2 an alternative view of the confluent 2-SUSY algorithm will be
elaborated. The restrictions imposed onto $u(x)$ in order to obtain
non-singular final potentials will be analysed in section 3. In section 4
we will apply the technique to the free particle, one-soliton well and
standard harmonic oscillator. 

\section{Second order supersymmetric quantum mechanics}

The second order supersymmetric quantum mechanics (2-SUSY QM) is a
particular realization of the standard supersymmetry algebra with two
generators \cite{ais93,aicd95,sa96,fe97,fgn98,ro98}: 
\begin{equation}
\{ Q_j,Q_k\} = \delta_{jk} H_{\rm ss}, \quad [H_{\rm ss},Q_j]=0, \quad
j,k=1,2, \label{susyal}
\end{equation}
where $Q_1 =(Q^\dagger + Q)/ \sqrt{2}, \ Q_2 = (Q^\dagger -
Q)/(i\sqrt{2})$,
\begin{eqnarray}
& Q = \left(\begin{matrix} 0 & A \\ 0 & 0 \end{matrix}
\right), \quad Q^\dagger = \left(\begin{matrix} 0 & 0 \\ A^\dagger & 0
\end{matrix}
\right), \\
& H_{\rm ss} =  
\left(\begin{matrix} A A^\dagger & 0 \cr 0 & A^\dagger A
\end{matrix} \right) = 
\left(\begin{matrix} (\wt H-\epsilon_1)(\wt H-\epsilon_2) & 0 \cr 0 & 
(H-\epsilon_1)(H-\epsilon_2)
\end{matrix} \right),
\label{factorized}
\end{eqnarray}
and $H, \ \wt H$ are two intertwined Schr\"odinger Hamiltonians: 
\begin{eqnarray}
& \wt HA = AH,  \label{intertwining} \\
& \wt H=-\frac{d^2}{dx^2}+\wt V(x), \quad H=-\frac{d^2}{dx^2}+V(x),
\label{horiginal} \\
& A=\frac{d^2}{dx^2}+\eta(x)\frac{d}{dx}+\g(x)\,.
\label{defA_seg_ord}
\end{eqnarray}
The relations between $\eta(x), \ \gamma(x), \ V(x)$ and $\wt V(x)$, are: 
\ba
& \wt V = V + 2 \eta^\prime, \label{Vt_V_eta_prime_a1}\\
& \g = d - V+\eta^2/2 - \eta^\prime/2\,,
\label{gam_eta} \\
& \eta \eta^{\prime\prime}-(\eta^{\prime})^2/2
+\eta^2(\eta^2/2 - 2 \eta^{\prime}-2 V + 2d ) + 2 c=0\,,
\label{eq_eta}  
\ea
where, in terms of $c\in{\mathbb R}, \ d\in{\mathbb R}$, the factorization
energies read $\epsilon_1 = d + \sqrt{c}$, $\epsilon_2 = d - \sqrt{c}$.
Suppose that $V(x)$ is a given exactly solvable potential; then $\wt V(x)$
will be effectively determined if we find explicit solutions $\eta(x)$ to
the non-linear second order differential equation (\ref{eq_eta}). 
Depending on the sign of $c$, two essentially different cases arise. 

If $c\neq 0$ then $\epsilon_1\neq\epsilon_2$, and we must look for
solutions of the two Riccati equations: 
\ba
\b^\prime_i +\b_i^2=V-\e_i\,, \quad i=1,2. \label{eqRicbeta_p}
\ea
Having $\beta_1(x)$, $\beta_2(x)$, we get two different equations for
$\eta(x)$ (see, e.g., \cite{ro98,fmr02}) 
\ba
\eta^{\prime}&=&\eta^2+2 \b_1\eta+\e_2-\e_1\,, \label{eta_beta_p}\\
\eta^{\prime}&=&\eta^2+2 \b_2\eta+\e_1-\e_2\,. \label{eta_beta_m}   
\ea
By subtracting them, we arrive to a finite difference algorithm for
$\eta(x)$:
\begin{equation}
\eta(x)=(\e_1-\e_2)/(\b_1-\b_2)\,.
\label{eta_beta_eps}
\end{equation}

On the other hand, the {\it confluent} case arises for $c=0$ implying that
$\epsilon_1=\epsilon_2\equiv \epsilon = d$. In this situation, we look for
solutions to just one Riccati equation \cite{mnr00}
\begin{equation}
\b^\prime+\b^2=V-\epsilon\,,  \label{eqRicbeta_0}
\end{equation}
and $\eta(x)$ satisfies an equation arising when $\epsilon_1=\epsilon_2$
in (\ref{eta_beta_p},\ref{eta_beta_m}):
\begin{equation}
\eta^{\prime}=\eta^2 + 2 \beta\eta\,.\label{ricc_eta_Bern}
\end{equation}
This is the Bernoulli equation, whose general solution is given by:
\begin{equation}
\eta(x) = -w'(x)/w(x), \label{nw}
\end{equation}
where
\begin{equation}
w(x) = w_0 - \int e^{2\int\beta(x)dx}dx\,. \label{confluente}
\end{equation}

In the next section we will analyse the conditions which grant that the
confluent 2-SUSY transformations are non-singular. This provides the
simplest way of ensuring that, departing from an exactly solvable initial
potentials $V(x)$, we arrive as well at an exactly solvable regular $\wt
V(x)$ (see e.g. the discussion in \cite{mnn98}).

\section{Confluent non-singular transformations}

Let us express first the confluent formulae of section 2 in terms of
solutions of the initial Schr\"odinger equation obtained from
(\ref{eqRicbeta_0}) by the change $\beta(x) = u'(x)/u(x)$: 
\begin{equation}
-u''(x) + V(x) u(x) = \epsilon u(x). \label{schro}
\end{equation}
Thus, up to an unimportant constant factor (see (\ref{nw})), the key
function $w(x)$ becomes
\begin{equation}
w(x) = w_0 - \int_{x_0}^x u^2(y)dy, \label{wu}
\end{equation}
and the confluent 2-SUSY potential $\wt V(x)$ is given by:
\begin{equation}
\wt V(x) = V(x) - 2 [w'(x)/w(x)]'. \label{npw}
\end{equation}

It is clear now that in order to arrive at real non-singular potentials
$\wt V(x)$ we have to use real solutions $u(x)$ of (\ref{schro}) inducing
a nodeless $w(x)$. Let us notice that
\begin{equation}
w'(x) = - u^2(x),
\end{equation}
meaning that $w(x)$ is decreasing monotonic, so the simplest way of
avoiding its zeros is to look for the appropriate asymptotic behaviour for
$u(x)$. Two different situations are worth considering.

(i) Suppose first that $\epsilon = E_m$ is one of the discrete eigenvalues
of $H$ and the transformation function is the corresponding {\it
normalized} physical eigenfunction, $u(x) = \psi_m(x)$.  Denote by $\nu_+$
the following finite integral: 
\begin{equation}
\nu_+ \equiv \int_{x_0}^\infty u^2(y)dy. \label{bsgp}
\end{equation}
It is straightforward to show that:
\begin{equation}
\lim_{x\rightarrow -\infty} w(x) = w_0 - \nu_+ + 1 \equiv
\nu + 1
\end{equation}
and
\begin{equation}
\lim_{x\rightarrow +\infty} w(x) = \nu.
\end{equation}
It turns out that $w(x)$ is nodeless if either both limits are positive or
both negative, leading to the $\nu$-domain where the confluent 2-SUSY
transformation is non-singular:
\begin{equation} 
\nu \in {\mathbb R} \setminus (-1,0)=(-\infty,-1]\cup[0,\infty).
\label{wdomainsi}
\end{equation}

(ii) Suppose now that the transformation function $u(x)$ is a {\it
non-normalizable} solution of (\ref{schro}) associated to a real
factorization energy $\epsilon\not\in{\rm Sp}(H)$ such that
\begin{equation}
\lim_{x \rightarrow \infty}u(x)=0 \quad {\rm and} \quad
\nu_+ \equiv \int_{x_0}^\infty u^2(y) dy < \infty. \label{gammap}
\end{equation}
If this is the case we can show that: 
\begin{equation}
\lim_{x\rightarrow -\infty} w(x) = w_0 + \int_{-\infty}^{x_0} u^2(y)dy =
\infty
\end{equation}
and
\begin{equation}
\lim_{x\rightarrow +\infty} w(x) = w_0 - \nu_+ \equiv \nu.
\end{equation}
By comparing both limits and taking into account that $w(x)$ is decreasing
monotonic, it turns out that $w(x)$ is nodeless if
\begin{equation}
\nu \geq 0. \label{wdomainp}
\end{equation}

Let us notice that the same $\nu$-restriction holds in the case when
\begin{equation}
\lim_{x\rightarrow -\infty}u(x) = 0 \quad {\rm and} \quad \nu_- \equiv
\int_{-\infty}^{x_0} u^2(y)  dy<\infty, \label{gammam}
\end{equation}
though now $\nu \equiv -(w_0 + \nu_-)$. 

Once the regularity of the confluent 2-SUSY algorithm is assured, we
analyse the spectrum of $\wt H$. From the intertwining relationship
(\ref{intertwining}) and the factorizations in (\ref{factorized}) we
immediately obtain normalized eigenstates $\vert \wt\psi_n \rangle$ of
$\wt H$ provided that $\nu$ satisfies either (\ref{wdomainsi})  in case
(i) or (\ref{wdomainp}) in case (ii) and $A \vert \psi_n \rangle \neq 0$:
\begin{equation}
\vert \wt\psi_n \rangle = (E_n-\epsilon)^{-1} A \vert \psi_n \rangle,
\label{psintilde}
\end{equation}
(so in the case (i) we cannot obtain $\vert \wt\psi_m \rangle$ of
(\ref{psintilde}) because $A\vert \psi_m \rangle = 0$). The orthonormal
set $\{\vert \wt\psi_n \rangle, n=0,1,2,\dots\}$ so constructed is not
automatically complete (we have to analyse yet the existence or not of an
extra normalizable function $\wt \psi$ belonging to the Kernel of
$A^\dagger$ which is orthogonal to all the $\vert \wt\psi_n \rangle,
n=0,1,2,\dots$).  To find $\wt \psi$ explicitly, let us factorize
$A^\dagger$ as follows:
\begin{eqnarray}
& A^\dagger = \left[\frac{d}{dx} + \beta(x)  \right] \left[\frac{d}{dx} -
\beta(x) - \eta(x) \right]. \label{adfactorized}
\end{eqnarray}
It turns out that the $\wt \psi\in{\rm Ker}(A^\dagger)$ we are looking for
is annihilated by the second factor operator of (\ref{adfactorized}):
\begin{equation}
\wt\psi(x) = n_0 e^{\int[\beta(x) + \eta(x)]dx} = n_0 u(x)/w(x),
\end{equation}
$n_0$ is a constant. It is straightforward to check that $\wt \psi(x)$ is
a normalized eigenfunction of $\wt H$ with eigenvalue $\epsilon$ for
$\nu\in{\mathbb R}\setminus [-1,0]$ in the case (i) with
$n_0=\sqrt{\nu(\nu+1)}$ and for $\nu>0$ in the case (ii) with
$n_0=\sqrt{\nu}$. On the other hand, $\wt \psi(x)$ becomes
non-normalizable for $\nu=-1,0$ in the case (i) or for $\nu=0$ in the case
(ii). Thus, when $\epsilon=E_m$ and $u(x)=\psi_m(x)$ it turns out that
${\rm Sp}(\wt H) = {\rm Sp}(H)$ if $\nu\in {\mathbb R}\setminus [-1,0]$
while the level $E_m$ is not present in ${\rm Sp}(\wt H)$ for $\nu=-1,0$
(in this case $E_m$ has been `deleted' in order to generate $\wt H$). On
the other hand, when $\epsilon\not\in{\rm Sp}(H)$ and $u(x)$ obeys either
(\ref{gammap}) or (\ref{gammam}) it turns out that ${\rm Sp}(\wt H) =
\{\epsilon\}\cup{\rm Sp}(H)$ for $\nu>0$ but ${\rm Sp}(\wt H) = {\rm
Sp}(H)$ for $\nu=0$.  For all the other $\nu$-values ($\nu\in(-1,0)$ in
case (i) and $\nu<0$ in case (ii)) it arises a singularity in $\wt V(x)$
due to the existence of a zero in $w(x)$. We notice, in particular, that
the case (ii) allows to generate a single level above the ground state
energy of $H$, a mechanism which cannot be directly implemented in the
1-SUSY treatment.

Let us remark that our confluent 2-SUSY procedure coincides with the
Abraham-Moses generation technique of creating, deleting or changing the
normalization of a single energy level \cite{am80} (see also
\cite{ni84,su85}). The same procedure, known as binary Darboux
transformations \cite{ms91}, has been employed to generate bound states
embedded in the continuum \cite{zs90,psp93,st95}. 

\section{The simplest applications}

Let us analyse now the simplest applications of the confluent 2-SUSY
algorithm. 

(a) Consider first the free particle for which $V(x)=0$. For a fixed
arbitrary $\epsilon <0$ which does not belong to ${\rm Sp}(H)$ there are
two asymptotically vanishing transformation functions: 
\begin{equation}
u(x)= \sqrt{2k} e^{\pm kx}, \quad \epsilon = -k^2.
\end{equation}
A direct calculation leads to:
\begin{equation}
w(x) = \mp 2e^{\pm k(x+x_1)} \cosh[k(x-x_1)],
\end{equation}
where $\nu = e^{\pm 2 k x_1} >0$. By substituting those expressions
into (\ref{npw}) we obtain the P\"oschl-Teller potential in both cases:
\begin{equation}
\wt V(x) = -2 k^2 {\rm sech}^2[k(x-x_1)],
\end{equation}
which has a bound state at $\epsilon = -k^2$.

(b) Take now the previous P\"oschl-Teller as the initial potential: 
\begin{equation}
V(x) = -2 k_0^2 {\rm sech}^2(k_0 x),
\end{equation}
and denote the ground state energy as usual, $E_0 = - k_0^2$. 

Let us consider first the case when $\epsilon = E_0$ and $u(x)$ is the
normalized ground state: 
 \begin{eqnarray}
& u(x) = \sqrt{\frac{k_0}2} \, {\rm sech}(k_0 x).
\end{eqnarray}
A straightforward calculation leads to: 
\begin{eqnarray}
& w(x) = \nu + \frac12 - \frac12 \tanh(k_0x), 
\end{eqnarray}
which produces once again the P\"oschl-Teller potential:
\begin{equation}
\wt V(x) = -2 k_0^2 {\rm sech}^2k_0(x-x_1),
\end{equation}
where $\tanh(k_0 x_1) = 1/(1+2 \nu)$.

Suppose now that $\epsilon =-k^2 \neq E_0, \ k\in{\mathbb R}$.  The
solutions with the right asymptotic behaviour are here: 
\begin{equation}
u(x) = \sqrt{2k} e^{\pm kx} [k_0 \tanh (k_0 x) \mp k],
\end{equation}
leading to:
\begin{equation}
w(x) = \mp \{\nu + e^{\pm2 k x}[k^2 + k_0^2 \mp2 k k_0 \tanh(k_0 x)] 
\}. \label{goodwpm}
\end{equation}
It turns out that the confluent 2-SUSY potential $\wt V(x)$ acquires the
Bargmann form:
\begin{equation}
\wt V(x) = -\frac{2(k_2^2-k_1^2)[k_1^2 {\rm sech}^2k_1(x-x_1) + k_2^2
{\rm csch}^2k_2(x-x_2)]}{[k_1\tanh k_1(x-x_1) - k_2\coth k_2(x-x_2)]^2}, 
\end{equation}
where for $k>k_0$ we need to take $k_1 = k_0, \ k_2 = k, \ \nu = (k^2 -
k_0^2)e^{\pm 2 k x_2},$ $e^{\pm 2k_0x_1} = (k+k_0)/(k-k_0)$ while for
$k<k_0$ we require $k_1 = k, \ k_2 = k_0, \ \nu = (k_0^2 - k^2)e^{\pm 2
k x_1},$ $e^{\pm 2k_0x_2} = (k+k_0)/(k_0-k)$. 

(c) Finally, let us analyse the harmonic oscillator potential: 
\begin{equation}
V(x) = x^2,
\end{equation}
which has a purely discrete spectrum composed of $E_n = 2 n + 1, \
n=0,1,\dots$ and eigenfunctions given by: 
\begin{equation}
\psi_n(x) = (\sqrt{\pi}2^n n!)^{-1/2} e^{-x^2/2} H_n(x), \quad n=0,1,\dots
\label{eigenfosc}
\end{equation}
where $H_n(x)$ are the Hermite polynomials. 

Let us suppose first that $\epsilon = E_m$ with $m$ fixed, $u(x)$ being
the corresponding normalized eigenfunction $\psi_m(x)$ of
(\ref{eigenfosc}). The calculation of (\ref{wu}) with $x_0=0$ leads to: 
\begin{eqnarray}
\hskip-0.5cm& w(x) \!=\! \nu + \frac12 -
x
\begin{matrix}
{}_{m_0} \\
\sum \\ 
{}^{s=0} 
\end{matrix}
\frac{(-1)^{m_0 + s}\Gamma(m_1)
(2x)^{2m_1-2s-1}}{2^{\delta+1}\sqrt\pi(m_1-s)\Gamma(\delta + \frac12) 
(m - 2s)!s!}\,{}_2F_2(m_1,m_1-s; \delta
+\frac12,m_1 +1-s; -x^2) \label{wods} 
\end{eqnarray}
where ${}_2F_2(a_1,a_2;b_1,b_2;z)$ is a generalized Hypergeometric
function \cite{er53}, $m_0 = (m-\delta)/2, \ m_1 = (m+\delta +1)/2$,
$\delta=0$ if $m$ is even but $\delta = 1$ if $m$ is odd. It turns out
that $\wt V(x)$ is isospectral to the oscillator potential, a case
illustrated in figure 1 for $\epsilon = 7$ ($m=3$) and $\nu = -5/4$.  Let
us notice the already involved explicit expression of $w(x)$ in
(\ref{wods}). 

\begin{figure}[ht]
\centering \epsfig{file=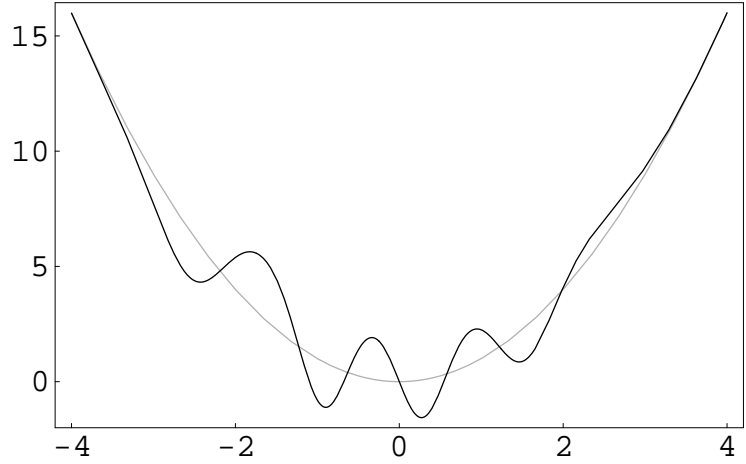, width=10cm}
\caption{\footnotesize The confluent 2-SUSY partner potential (black
curve) isospectral to the oscillator (gray curve) generated by employing
the normalized eigenfunction (\ref{eigenfosc}) for $n=3, \ \epsilon = E_3
= 7$ and $\nu = -5/4$.}
\end{figure}

In turn, for $\epsilon\not\in{\rm Sp}(H)$ the asymptotically vanishing
Schr\"odinger solutions become: 
\begin{eqnarray}
& u(x) = 
e^{-\frac{x^2}2}\big[{}_1\!F_1(\frac{1-\epsilon}4,\frac12;x^2)
\pm 2 x\frac{\Gamma(\frac{3-\epsilon}4)}{\Gamma(\frac{1-\epsilon}4)}
\, {}_1\!F_1(\frac{3-\epsilon}4,\frac32;x^2)\big], \label{solpm}
\end{eqnarray}
where ${}_1F_1(a,c;z)$ is the Kummer hypergeometric series. The explicit
expression for $w(x)$ is too involved to be shown here (three infinite
sums of kind (\ref{wods}) arise in this case). Alternatively, we performed
a numeric calculation of $\wt V(x)$ for $\epsilon = 8$ taking the
solution $u(x)$ of (\ref{solpm}) with the upper plus sign and $w_0=-5$,
$x_0=0$ in (\ref{wu}) (see figure 2). The spectrum of $\wt V(x)$ is
composed of the oscillator eigenenergies $E_n=2n+1, \ n=0,1,\dots$ plus a
new level at $\epsilon=8$.  This illustrates clearly the possibility
offered by the confluent 2-SUSY algorithm of creating one single level
above the ground state energy of $H$.

\begin{figure}[ht]
\centering \epsfig{file=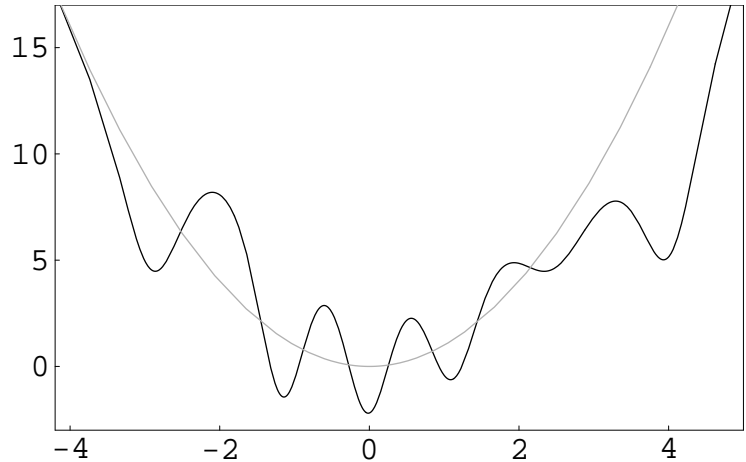, width=10cm}
\caption{\footnotesize The confluent 2-SUSY partner potential $\wt V(x)$
(black curve) of the oscillator (gray curve) generated by employing the
Sch\"odinger solution (\ref{solpm}) with the upper $+$ sign and
$\epsilon=8, \ w_0=-5, \ x_0=0$. The potential $\wt V(x)$ has an extra
bound state at $\epsilon=8$ compared with the oscillator spectrum.}
\end{figure}

We conclude that the second order supersymmetric quantum mechanics is a
powerful tool for designing in a simple way potentials with given spectra,
a subject supplying us of solvable models with possible applications in
the physical sciences (see e.g. \cite{dr97}).

{\small
{\bf Acknowledgements.} 
The support of CONACYT (M\'exico) is acknowledged. One of the authors
(ESH) also acknowledges the financial support of the {\it Instituto
Polit\'ecnico Nacional}.}

\end{document}